\documentclass{PoS}
\usepackage{amsmath}

\title{Hamiltonian Analysis of Asymptotically Safe Gravity}

\ShortTitle{Hamiltonian Analysis of Asymptotically Safe Gravity }

\author{\speaker{Gabriele Gionti, S.J.}\\
        Specola Vaticana, V-00120 Vatican City, Vatican City State and\\
Vatican Observatory Research Group Steward Observatory, The University Of Arizona,\\ 
933 North Cherry Avenue, Tucson, Arizona 85721, USA.\\
INFN, Laboratori Nazionali di Frascati, Via E. Fermi 40, 00044 Frascati, Italy.\\
         E-mail: \email{ggionti@specola.va}}

\abstract{Recent results based on renormalization group approaches to Quantum Gravity suggest that the Newton's and cosmological constants  should be treated as dynamical variables whose evolution depend on the characteristic energy scale of the system. An open question is how to embed this modified Einstein's theory in the Dirac constrained dynamics. In this work, the Hamiltonian formalism for a renormalization-group scale dependent Einstein-Hilbert action is discussed  paying particular attention to Dirac's constraint analysis. It is shown that the algebra of the Dirac's constraints, in some cases, is closed. Applications to the physics of the Early Universe are explicitly discussed assuming the framework of Asymptotic Safety. In particular, it is shown that in the minisuperspace with FLRW metric, RG-improved Friedmann equations have bouncing and emergent Universes solutions also for flat and negative curvature.}

\FullConference{Corfu Summer Institute 2017 'School and Workshops on Elementary Particle Physics and Gravity'\\
		2-28 September 2017\\
		Corfu, Greece}

\begin{document}
\section{Introduction}
It is well known that classical General Relativity is a quite successful phenomenological theory at laboratory, solar system, galactic and extragalactic scales and in general for length scales $l \gg  l_{Pl}\approx 10^{-33}cm$, where $l_{Pl}$ is the Planck length. Singularity problems of Einstein's equations at Planck length and the quantum behaviour of matter and energy at small distances (high energy) suggest that a quantum version of the gravitational field (Quantum Gravity) should be found. There are many different approaches to Quantum Gravity: String Theory, Loop Quantum Gravity, Non-Commutative Geometry, Causal Dynamical Triangulations, Poset Theory, Asymptotic Safety etc. 

As the Newton's constant has a negative mass dimension, the perturbative quantization of General Relativity leads to a (perturbative) non-renormalizable theory. In general,  perturbative non-renormalizable theories  have a number of counter terms which increase as the loop orders. This implies that the renormalization process introduces infinitely many parameters so that the resulting theory does not have any predictive power \cite{lauscherreuter2001}.  This is not a dead end, because a perturbatively non-renormalizable theory might be renormalizable under a generalized notion of renormalizability based on non-perturbative arguments. This non-perturbative renormalizability, introduced by K. Wilson  \cite{wilsonfourQFT}, is related to the existence of a Non-Gaussian Fixed Point (NGFP) which guarantee the finiteness of the theory in the ultraviolet limit\cite{NiederReuter}.

The {\it Asymptotics Safety} conjecture dates back to Weinberg \cite{1979weinberg}. He suggested that General Relativity might be a non-perturbatively renormalizable Quantum Field Theory if the gravitational RG-flow approaches a non-trivial fixed point in the high energy limit. He himself proved that NGFP exists in 2+$\epsilon$ dimensions \cite{1979weinberg}. In d=4 a NGFP exists in the case of {Einstein-Hilbert} truncation \cite{2001ReuterSaueressig}. The main idea of this approach is that if one has a classical action of Gravity, in the Riemannian case, coupled with $a_{i}$ constants coupled to $O^{i}(x,g)$ operators, $x$ and $g$ being, respectively, the space-time coordinates and the metric tensor $g$ \cite{Guarnieriphd},

\begin{equation}
S(M,g)=\int_{M}d^{4}x\sqrt{g}\sum^{\infty}_{i=0}a_{i}O^{i}(x,g)\;\;\;\;, 
\label{gen}
\end{equation}

$M$ is the four dimensional differentiable manifold. The renormalizable group is defined once one fixes an infrared cutoff $k$ and writes the renormalization group equations in terms of the dimensionless coupling constants $\tilde{a}_{i}(k)$ and the $\beta$-functions in the following manner \cite{Guarnieriphd}

\begin{equation}
k\partial_{k}\tilde{a}_{i}(k)=\beta_{i}(\tilde{a}_{1}(k), \tilde{a}_{2}(k), \tilde{a}_{3}(k)...)
\label{RG}
\end{equation}

A point $\tilde{a}_{\star}$ is a NGFP if it is a non trivial zero of the beta-functions, that is $\beta_{i}(\tilde{a}_{\star})=0 \; \forall i$  and $\tilde{a}_{\star}\neq 0$. 

Once one has found a the NGFP, the next step is to linearize previous equation \cite{Guarnieriphd}

\begin{equation}
k\partial_{k}\tilde{a}_{i}(k)=\sum_{j}B_{ij}\left(\tilde{a}_{j}(k)-\tilde{a}_{\star j}(k)\right)\;\;, 
\label{lirg}
\end{equation}

where one has assumed the following definitions: 

\begin{equation}
B_{ij}\equiv\partial_{j}\beta_{i}(\tilde{a}_{\star})\;\;\;\;,\;\;\;\;B=\left(B_{ij}\right)\;\;\;\;.
\label{pos}
\end{equation}

The general solution to the previous linear equation can be written in the following form

\begin{equation}
\tilde{a}_{i}(k)=a_{\star i} +\sum_{I}C_{I}V^{I}_{i}\left( \frac{k_{0}}{k} \right)^{\Theta _{I}}\;\;\;\;,
\label{gena}
\end{equation}

where $V^{I}$ are right-eigenvectors, solutions of the eigenvalue equation (matrix equation)

\begin{equation}
B\;V^{I}=-\Theta_{I}V^{I}
\label{eigen}
\end{equation}

$\Theta_{I}$ being the critical exponents. Now, notice that the fact that one assumes $\tilde{a}_{i}(k) \longmapsto a_{\star i}$ when $k\mapsto \infty$ requires that $C_{I}=0 \forall I$ in case 
$\mathrm{Re} \;\Theta_{I} <0$. The Ultra-Violet(UV) critical surface $S_{UV}$ is defined as the number of independent renormalization group trajectories hitting the fixed point as $k\mapsto \infty$. The dimension $\Delta_{UV}$ of this surface is the dimension of $S_{UV}$. Said in another way, the dimension of the critical surface is the number of independent attractive directions or, equivalentely,  the number of eigenvalues $\Theta$ with $\mathrm{Re} \;\Theta_{I} >0$. The resulting quantum theory has $\Delta_{UV}$ free parameters. If this number is finite, then the theory is predictive as a pertinent renormalizable model with $\Delta_{UV}$ renormalizable couplings. 

These considerations hold, in general, but has been introduced for the perturbative renormalization group (RG). In the non perturbative case one starts from a Wilson-type, coarse-grained, free energy functional 

\begin{equation}
\Gamma_{k}\left[g_{\mu\nu}\right]\;\;\;\;,
\label{fe}
\end{equation}

where $k$ is the infrared cut-off. $\Gamma_{k}$  contains all the quantum fluctuations with momenta $p>k$ and not yet of those with $p<k$. The modes $p<k$ are suppressed in the path-integral by a mass-square  type term $R_{k}(p^{2})$.

The behavior of the free-energy functional interpolates between $\Gamma_{k\mapsto \infty}=S$, $S$ being the classical (bare) action, and $\Gamma_{k\mapsto 0}=\Gamma$, $\Gamma$ being the standard effective action. $\Gamma_{k}$ satisfies the RG-equation, called also the Wetterich equation \cite{WetterichFRG}, 

\begin{equation}
k\partial_{k}\Gamma_{k}=\frac{1}{2}Tr\left[(\delta^2\Gamma_k + R_k)^{-1}k\partial_k R_k\right]
\label{Wett}
\end{equation}

In general, since this $RG$-equation is quite complicate, one adopts a powerlul non perturbative approximation scheme: truncate the space of the action functional and project the RG flow onto a finite dimensional space. That is to say, one consider that the free energy functional $\Gamma_{k}$, formally, can be expanded in the following way

\begin{equation}
\Gamma_{k}[\cdot]=\sum_{i=0}^{N}g_{i}(k)k^{d_i}I_i[\cdot]\;\;\;\;,
\label{svilop}
\end{equation}

where $I_{i}[\cdot]$ are given "local or non local functionals" of the fields and $g_{i}(k)$. In the case of gravity, the following truncation ansatz is usually made:

\begin{equation}
I_{0}[g]=\int d^{4}x\sqrt{g}\;\;\;\;,I_{1}[g]=\int d^{4}x\sqrt{g}R\;\;\;\;,I_{2}[g]=\int d^{4}x\sqrt{g}R^{2}\;\;\;\;,\mathrm{etc.}
\label{trunca}
\end{equation}

The simplest truncation is the Einstein-Hilbert truncation which looks like

\begin{equation}
\Gamma_{k}=-\frac{1}{16\pi G_{k}} \int d^{4}x  \left( R-2 \bar{\lambda}_{k} \right) +\mathrm{g.f. } +\mathrm{g.t.}\;\;\;\,, 
\label{truncaEH}
\end{equation}

here g.f. means classical gauge fixing terms, while g.h. are ghost terms. There are two running parameters $G_{k}$, the Newton constant, which can be written in a dimensionless way as   $g(k)=k^{2}G_{k}$. In the same manner, the cosmological constant $\bar{\lambda}_{k}$ becomes $\lambda(k)=\bar{\lambda}_{k}/k^{2}$.

Inserting this ansatz into the flow (Wetterich) equation, one obtains "a projection" onto e finite dimensional space  \cite{NiederReuter}
\begin{equation}
Tr[...]=(...)\int \sqrt{g} +(...)\int \sqrt{g}R +...\;\;\;\;, 
\label{expo}
\end{equation}

and then the following finite-dimensional RG equations 

\begin{eqnarray}
k\partial_{k}g(k)=\beta_{g}(g,\lambda)\\ \nonumber
k\partial_{k}\lambda(k)=\beta_{\lambda}(g,\lambda)\;.
\label{finite}
\end{eqnarray}

The solutions of this equations provide the scaling relation for the a-dimensional gravitational constant $g(k)$ and the a-dimensional cosmological constant $\lambda(k)$. 

\section{Modified Einstein-Hilbert Action and Lorentzian ADM Asymptotic Safe Gravity}

It is an old idea, which dates back to Dirac \cite{1972Dirac},  to consider that the gravitational constant $G$ is not really a constant but varies as function of the Space-Time coordinates, $G=G(x)$ \cite{BransDicke} The first idea in this direction dates back to Brans and Dicke theory \cite{BransDicke}, who proposed a coupling of gravity with a scalar field $\phi(x)$ of the type $\phi(x)\equiv 1/G(x)$ \cite{ReuterWeyer}.

The method proposed here is quite different respect to the usual Brans-Dicke theory. In fact, the scalar field $\phi(x)$ is a true dynamical variable in Brans-Dicke theory with a kinetic term, whose equation of motion is determined by varying the action with respect to the field $\phi(x)$ \cite{ReuterWeyer}. Instead, here, one aims to look for a modified theory of General Relativity. In fact, following the general guide-line of Asymptotic Safety approach to Quantum Gravity, the first step is to find the $k$-dependence of the coupling constants, in our case $G$ and $\Lambda$, by the Renormalization Group approach as explained just above. The second step is to fix the dependence from space-time $x$ of the infrared cutoff $k$, that is $k=k(x)$. This identification is generally made on the base of either symmetrical or physical arguments. Therefore $G(k(x))=G(x)$ and $\Lambda(k(x))=\Lambda(x)$ become Space-Time functions that cannot be determined on the base of a Lagrangian dynamic \cite{ReuterWeyer}. Therefore they behave, technically, either as  external or, equivalently, as non-geometrical fields. Then the variation of the modified Einstein-Hilbert Lagrangian under the variation of the metric tensor $g$ does not affect $G(x)$ and $\Lambda(x)$ which should be considered given functions. Reuter and Weyer \cite{ReuterWeyer} remark that the modified Einstein Equations, derived from modified Einstein Hilbert Lagrangian, should contain some extra integrability conditions which should put constraints on $G(x)$ and $\Lambda (x)$, or further constraints on the cutoff identification $k(x)$. Reuter and Weyer \cite{ReuterWeyer} start from the following modified Einstein-Hilbert action

\begin{equation}
S_{mEH}[g,G(x),\Lambda(x)]\equiv \frac{1}{16\pi}\int d^{4}x\sqrt{-g}\left(\frac{R}{G(x)}-2\frac{\Lambda(x)}{G(x)}\right)\;\;\;\;.
\label{mEH}
\end{equation}

Following \cite{Manrique}, one starts from a Lorentzian Metric (M,g) and consider an ADM metric decomposition  

\begin{equation}
g=-(N^{2}-N_{i}N^{i})dt \otimes dt +N_{i}(dx^{i} \otimes dt
+dt \otimes dx^{i})+h_{ij}dx^{i} \otimes dx^{j}\;\;\;\;, 
\label{metricADM}
\end{equation}

where $N=N(x)$ is a function on the four dimensional space-time and it is called "lapse", $N^{i}(x)$ are called "shifts", and $h_{ij}(x)$ is the three-metric on the space-like surfaces $\Sigma$ of the ADM foliation \cite{menotti}.  In this context, the regulator $R_{k}$ depends on the Laplacian on the three-dimensional spatial surfaces $\Sigma$. The infrared cut-off of the RG transformations is built from the spectrum of the Laplacian operator defined on the three-dimensional surfaces $\Sigma$. 

\section{ADM Analysis of Modified Einstein-Hilbert Lagrangian}

One consider, from now on, a Space-Time $(M,g)$ which is such that $M\equiv \Re \times \Sigma $, $\Re$ being the time-like direction, and $\Sigma$ the Space-like three-dimesional surface. The metric tensor $g$ inherits ADM decomposition form given by (\eqref{metricADM}). The extrinsic curvature term $K_{ij}$ is defined on the three-dimesional surface $\Sigma$ and has the following definition 

\begin{equation}
K_{ij}=\frac{1}{2}(-\frac{\partial h_{ij}}{\partial t}+{\bar \nabla}_{i}N_{j}+{\bar \nabla}_{j}N_{i})
\label{curvextrin}
\end{equation}

where the covariant derivative $\bar \nabla$ is a covariant derivative defined on the three-dimensional spatial surfaces $\Sigma$ through the three-dimensional spatial metric $h_{ij}$. The four-dimensional trace of the Ricci tensor ${}^{4}R$ is decomposed into ADM foliation in the following way \cite{AlfioGiampieroRubano}

\begin{equation}  
\sqrt{-g}{}^{4}R=N\sqrt{h}\left(K_{ij}K^{ij} - K^{2} + {}^{3}R\right) -2 \left(K \sqrt{h} \right),_{\;0} + 2 f^{i},_{i}\;\;\;\;\;, 
\label{scompongo}
\end{equation}

where 

\begin{equation}
f^{i}\equiv \sqrt{h} \left(K N^{i} - h^{ij} N,_{j}\right)\;\;\;\;. 
\label{bondo}
\end{equation}

It is useful to have in mind the following identities \cite{AlfioGiampieroRubano}

\begin{eqnarray}
{1 \over G}\left(K\sqrt{h}\right),_{0}&=&{G,_{0} \over G^{2}}K\sqrt{h}+\left(K\sqrt{h} \over G \right)_{,0} \label{idento1}\\ 
\
{1 \over G}{\partial f^{i} \over \partial x^{i}}&=&{G_{,i} \over G^{2}}f^{i} + {\partial \over \partial x^{i}} \left(f^{i} \over G \right)
\label{idento}
\end{eqnarray}

Once introduced these definitions, it is quite straightforward to write down the Einstein-Hilbert action into ADM coordinates $S_{ADM}(h_{ij}, N, N^{i})$ with the York boundary term. The latter is needed to make the variation of the Einstein-Hilbert action linear under the variations of the metric tensor \cite{menotti}

\begin{equation}
S_{ADM}[h_{ij},N, N^{i}]=\frac{1}{16\pi}\int_{R \times \Sigma}dt d^{3}x \sqrt{h}N \frac{1}{G(t,x)}\left({}^{4}R-2 \Lambda(t,x)\right)+\frac{1}{8\pi} \int_{\partial M}{K \sqrt{h}\over G(t,x)}d^{3}x\;\;.
\label{ADMnormal}
\end{equation}

This action can be simplified a lot if one uses the identities above \eqref{idento1}\eqref{idento} and suppose that $\Sigma$ is a closed manifolds (so that the total spatial divergence of  $f^{i} \over G$ resulting from \eqref{scompongo} and \eqref{idento}  yields zero contribution, having taken $\partial \Sigma=\emptyset$). So one gets, finally,

\begin{equation}
S_{ADM}(h_{ij},N, N^{i})={1\over 16\pi}\int_{R \times\Sigma} \left[{N \sqrt{h}\over G}(K_{ij}K^{ij}
-K^{2}+{ }^{(3)}R-2 \Lambda)
-2{G_{,0}\over G^{2}}K \sqrt{h}
+2{G_{,i}f^{i}\over G^{2}}\right]dtd^{3}x\;\;\;\;.
\label{extendoADM}
\end{equation}

 Until now one has considered this theory in the Lagrangian formalism. One wants to pass from the Lagrangian formalism with the variables $(N, N^{i}, h_{ij})$ to the Hamiltonian formalism in which there are positions and momenta coordinates. Therefore the first step, in this process, is to define the Lagrangian density ${\cal L}_{ADM}$ from previous equation (\ref{extendoADM}):
  
 \begin{equation}
 {\cal L}_{ADM}\equiv {1\over 16\pi} \left[{N \sqrt{h}\over G}(K_{ij}K^{ij}
-K^{2}+{ }^{(3)}R-2 \Lambda)
-2{G_{,0}\over G^{2}}K \sqrt{h}
+2{G_{,i}f^{i}\over G^{2}}\right]\;\;\;\;.
\label{densiLagr}
\end{equation}
 
From the previous definition, one gets the "spatial momenta" $\pi_{ij}$ 

 \begin{equation}
{\pi}^{ij}={\partial{{\cal L}_{ADM}} \over \partial{\dot h}_{ij}}= 
-\frac{\sqrt h}{16\pi G}\left(K^{ij}-h^{ij}K\right) +\frac{{\sqrt h}\;h^{ij}}{16\pi N G^{2}}\left(G_{,0}-G_{,k}N^{k}\right)
\label{momenta}
\end{equation}

Of course, it is straightforward to notice that these are momentum densities rather than momenta. Their integration, on the three-dimensional surfaces $\Sigma$, gives the momenta \cite{menotti}. The formula above \eqref{momenta} does not look very encouraging since there is not a direct relation between the momentum densities and the extrinsic curvature as in standard ADM \cite{menotti}. But one can define the following new variable  ${\tilde \pi}_{ij}$ in the following way

\begin{equation}
{\tilde \pi}^{ij}=\pi^{ij}-\frac{{\sqrt h}\;h^{ij}}{16\pi N G^{2}}\left(G_{,0}-G_{,k}N^{k}\right)
=-\frac{\sqrt h}{16\pi G}\left(K^{ij}-h^{ij}K\right)
\label{newmomnta}
\end{equation}

which have the usual relation with the extrinsic curvature $K_{ij}$ \cite{espositolibro}. Of course, one can define the conjugate momenta $\pi$ and $\pi_{i}$ to the lapse function $N$ and to the shift functions $N^{i}$. Following Dirac's constraint theory, they are zero on the constraint surface and are called primary constraints \cite{menotti}

\begin{equation}
 \pi=\frac{\partial {\cal L}_{ADM}}{\partial \dot {N}}\approx 0\;\;\;\;\;\;\pi_i=\frac{\partial {\cal L}_{ADM}}{\partial \dot {N}^i}\approx 0 
 \label{vincprima}
 \end{equation}
 
 Then one may reasonably ask if the following change of variables  
 
\begin{equation}
\left (N, N^{i}, h_{ij},\pi, \pi_{i}, \pi^{ij}\right)\mapsto \left (N, N^{i}, h_{ij}, \pi, \pi^{i}, {\tilde \pi}^{ij}\right)
\label{cambio}
\end{equation}

is a canonical transformation in the Hamiltonian formalism \cite{Arnoldlibro}. A change of variables of the kind (\ref{cambio}) is canonical if the symplectic two form $\Omega=dq^{i} \wedge dp_{i}$ is invariant under this transformation. Applying trivially this definition, a change of variable $(q^i, p_i) \mapsto (Q^i, P_i)$ is a canonical transformation if the following conditions hold \cite{Arnoldlibro}

\begin{eqnarray}
F\equiv \frac{\partial (q^1,...q^n,p_1,...,p_n)}{\partial (Q^1,...,Q^n,P_1,...,P_n)}&\;\;\;\; &
J\equiv
\begin{pmatrix} 
0 & I \\
-I & 0  
\end{pmatrix}\\ 
&F^{T}JF=J&
\label{canonical}
\end{eqnarray}

where $F$ is the Jacobian of the transformation of variables. It is a straightforward calculation to check that the new definition of the momenta \eqref{newmomnta} defines a change of variables \eqref{cambio} satisfying the requirement  \eqref{canonical}.

One is entitled to define the "canonical" Hamiltonian density ${\cal H}_{ADM}$ \cite{espositolibro} on the manifold fixed by the primary constraints \eqref{vincprima},

\begin{equation}
 {\cal H}_{ADM}={\pi}^{ab}{\dot h}_{ab}-{\cal L}_{ADM}\;\;\;\;,
 \label{cano}
 \end{equation}

from this Hamiltonian density, one can define the Hamiltonian density in the new canonical coordinates \eqref{cambio} through substitution \eqref{newmomnta} \cite{Arnoldlibro}.  

So implementing these transformations one gets

\begin{eqnarray}
{\cal H}_{ADM}&=&N\left((16\pi G)G_{abcd}{\tilde \pi}^{ab}{\tilde \pi}^{cd}-\frac{{\sqrt h}({}^{3}R-2\Lambda)}{16\pi G}\right)+2{\tilde \pi}^{ab}{\bar \nabla}_{a}N_{b}\\ \nonumber &+&
\frac{{\sqrt h}(G_{,0} -G_{,k} N^k){\bar \nabla}_{a}N^{a}}{8\pi G^2 N}
+\frac{G_{,i}{\sqrt h}h^{ij}}{8\pi G^{2}}N_{,j}
\label{hamiltonianagrande}  
\end{eqnarray}

in which $G_{abcd}$ is DeWitt supermetric \cite{espositolibro} defined in the following way 

\begin{equation}
G_{abcd}=\frac{1}{2 \sqrt h}\left (h_{ac}h_{bd}+h_{ad}h_{bc}-h_{ab}h_{cd}\right)
\label{DeWittmetric}
\end{equation}

One is now in a position to define the total Hamiltonian $H_{T}$ as \cite{dirac1966}
\begin{equation}
H_T=\int_{\Sigma}\left(\lambda \pi + \lambda^{i}\pi_{i}+{\cal H}_{ADM}\right)d^{3}x\;\;\;\;,
\label{hamilttotal}
\end {equation}

where $\lambda$ and $\lambda^{i}$ are Lagrange multipliers. Basic and well known considerations, integrations by parts and confront with standard 
Hamiltonian analysis of General Relativity \cite{menotti}, suggest to define the Hamiltonian constraint $\cal H$ and the momentum constraints ${\cal H}_i$ through the preservation of the primary constraints \eqref{vincprima}

\begin{equation}
{\cal H}\equiv \left\{\pi,H_T\right\}\;\;,\;\;{\cal H}_i\equiv \left\{\pi,H_T\right\}.
\label{defino}
\end{equation}

Recall that the Poisson brackets are defined as 

\begin{equation}
\{A,B\}=\int d^{3}x\left(\frac{\delta A}{\delta h^{ij}}\frac{\delta B}{\delta {\tilde\pi}_{ij}} - \frac{\delta A}{\delta {\tilde\pi}_{ij}}\frac{\delta B}{\delta h^{ij}},\right),
\label{parentesis}
\end{equation}

therefore the Hamiltonian constraint $\cal H$ results to be 

\begin{equation}
{\cal H}=(16\pi G)G_{abcd}{\tilde \pi}^{ab}{\tilde \pi}^{cd}-\frac{{\sqrt h}({}^{3}R-2\Lambda)}{16\pi G}-\frac{{\sqrt h}(G_{,0} -G_{,k} N^k){\bar \nabla}_{a}N^{a}}{8\pi G^2 N^2}-\nabla_{j}\left(\frac{G_{,i}{\sqrt h}h^{ij}}{8\pi G^{2}}\right)
\label{vinchamiltgrande}
\end{equation}

while the momenta constraints ${\cal H}_i$ are 

\begin{equation}
{\cal H}_{i}=-2{\bar \nabla}^{a}{\tilde \pi}_{ai}+\frac{{\sqrt h}(-G_{,i}){\bar \nabla}_{a}N^{a}}{8\pi G^2 N}-{\sqrt h}{\bar\nabla_{i}}\left(\frac{G,_{0}-G_{,k}N^k}{8\pi G^2 N}\right)\;\;\;\;.
\label{vincconstrgrande}
\end{equation}

The previous expressions of the Hamiltonian constraint $\cal H$ and the momentum constraints ${\cal H}_i$ appear quite complicated. 
The first check, one can do, is to see how these functions behave under gauge transformations. Following \cite{menotti} one can calculate the following (gauge) transformation on the three-dimensional spatial surfaces $\Sigma$ and gets

\begin{equation}
\{h_{ij},\int d^{3}x \tilde{N}^{i}{\cal{H}}_{i}\}={\cal{L}}_{\mathbf {\tilde{N}}} h_{ij}\;\;\;\;,
\label{tremetric}
\end{equation}
 
where ${\bf \tilde{N}}=\left( {\tilde{N}}^{i} \right)$ is a generic three-dimensional vector field on $\Sigma$. Therefore the momentum constraints ${\cal H}_i$ are still the generators of the gauge transformations on the three-dimensional metric $h_{ij}$. Following the same reasonings for the momenta  ${\tilde \pi}^{ij}$, one obtains

\begin{equation}
\left\{{\tilde \pi}^{ij},\int d^{3}x {\tilde N}^{i}{\cal{H}}_{i}\right\}=\int d^{3}x{\cal{L}}_{\bf {\tilde{N}}} {\tilde\pi}^{ij}+{\bar \nabla}_{a}\left[\frac{{\tilde N}^{s}}{2}\left(\frac{G_{,\;s}}{8\pi G^2 N}\right)N^{a}h^{ij}\sqrt{h}\right]\;\;\;\;, 
\label{momentatrespace}
\end{equation}

the momentum constraints ${\cal H}_{i}$ will be still the generators of the diffeomorphism transformations on $\Sigma$ for the momentum densities ${\tilde \pi}^{ij}$ if, sufficient condition, $G,_{s}=0$, that means $G=G(t)$. 
Taking into account the standard results of the Hamiltonian theory of Einstein General Relativity as regards the constraint algebra \cite{menotti} , one finds that the non-zero part of the Poisson brackets on the Hamiltonian constraint reduces to

\begin{equation}
\{\int d^{3}x{\tilde N}(x){\cal{H}}(x),\int d^{3}x'{\tilde N}'(x'){\cal{H}}(x')\}=\int d^3y({\tilde N}'{\bar\nabla}_{i}{\tilde N}-{\tilde N}{\bar\nabla}_{i}{\tilde N}')\frac{G_{,0}N^i}{GN^2}\tilde\pi\;.
\label{commutogrande}
\end{equation}
 
Clearly this says that in order to preserve the Hamiltonian constraint and then time-diffeomorphims, in the ADM splitting, one has to impose the sufficient condition $N^{a}\approx0$. Furthermore looking at the Hamiltonian constraint \eqref{vinchamiltgrande} and the momentum constraints \eqref{vincconstrgrande}, the only possibility they stay first class constraints is to impose strongly $N^{a}=0$ and $N=N(t)$. This implies the right ADM metric to start is not \eqref{metricADM} but one with reduced gauge invariances 

\begin{equation}
g=-N^{2}(t)dt \otimes dt+h_{ij}dx^{i} \otimes dx^{j}
\label{normalcoordinates}
\end{equation}

in which the shifts $N^{i}$ are put to zero and $N=N(t)$. This is  ADM metric in Gaussian normal coordinates \cite{Wiltshire}. The ADM-Hamiltonian density ${\cal H}_{ADM}$ reduces to 

\begin{equation} 
{\cal H}_{ADM}=N\left((16\pi G)G_{abcd}{\tilde \pi}^{ab}{\tilde \pi}^{cd}-\frac{{\sqrt h}({}^{(3)}R-2\Lambda)}{16\pi G}\right)\;\;\;\;,
\label{Hamilrido}
\end{equation}

One has only the primary constraint $\pi \approx 0$ and the Hamiltonian constraint ${\cal H}$ 

\begin{equation}
{\cal H}=\left((16\pi G)G_{abcd}{\tilde \pi}^{ab}{\tilde \pi}^{cd}-\frac{{\sqrt h}({}^{(3)}R-2\Lambda)}{16\pi G}\right)\;\;\;\;, 
\label{constraint}
\end{equation}
  
  and they are first class. The Hamiltonian constraint is preserved as in General Relativity \cite{menotti} and the Hamiltonian analysis does not impose any restriction on the functional form of $G(x)$. 

\section{Cosmologies of the Sub-Planck Era}
As a straightforward application of all previous considerations, one can study a cosmological minisuperspace model based on FLRW metric

\begin{equation}
ds^2 = -N(t)^2 dt^2 +\frac{a(t)^2}{1-K r^2} dr^2 +a(t)^2 (r^2 d\theta^2 + r^2\sin\theta d\phi^2)\;\;\;\;,
\label{FLRWmetric}
\end{equation}

where $N(t)$ is the Lapse function discussed in the considerations above, $a(t)$ is the scale factor of the universe and $K$ assumes values $-1, 0, 1$ depending on the topology one consider, that is respectively hyperbolic-open Universes, flat-infinite Universe or closed-elliptic  Universe. All the details on this particular cosmological case can be found in the reference \cite{AlfiomeAlessia}. Notice that FLRW metric is an ADM-metric in Gaussian normal coordinates \eqref{normalcoordinates} Following the discussion above, one considers a renormalization group modified Einstein-Hilbert action $S$ 

\begin{equation}
S =\int_{M} d^4 x \sqrt{-g} \, \left\{\frac{R-2\Lambda(k)}{16\pi G(k)} + \mathcal{L}_m\right\}+\frac{1}{8\pi} \int_{\partial M}{K \sqrt{h}\over G(k)}d^{3}x\;\;\;\;.
\label{actionpunto}
\end{equation}

This action, respect to formula (\ref{mEH}), does not contain the dependence of the infrared cut-off $k$ by the space-time coordinates $x$, $k=k(x)$. $G(k)$ is the gravitational constant and $\Lambda(k)$ the cosmological constant. They both depended from $k$. $M$ is a Lorentzian Manifold with boundary $\partial M$ and a York term has been added in the previous formula.  $\mathcal{L}_{m}$ is the the Lagrangian for the matter fields. 

The very fact of using a homogeneous and isotropic FLRW Universe implies that the infrared cutoff $k$, for symmetry reasons,  can depend only on cosmological time $t$, $k=k(t)$ and so this implies $G$ and $\Lambda$ are function of $t$ only

\begin{equation}
G\equiv G(k(t)),\;\;\;\; \Lambda\equiv \Lambda(k(t)). 
\label{equivo}
\end{equation}

In principle this dependence of $k$ from $t$ could be either explicit or implicit via the scale factor $a(t)$, $k=k(t,a(t),{\dot a}(t), {\ddot a}(t)...)$ \cite{AlfioGiampieroRubano}, \cite{2002PhRDAlfioReuter}\cite{2002PhLBAlfioReuter}. As already explained in section 2, recent work \cite{Manrique} has shown that for the ADM formalism \cite{1960JMPADM} \cite{1961PhRDADM} \cite{1960PhR1ADM} \cite{1960PhR2ADM} the infrared cutoff of the RG transformations is built from the spectrum of the Laplacian operator defined on the three-dimensional surfaces $\Sigma$. In particular $k\sim a^{-1}$ in the case of FLRW metric. 

Let us also assume that the matter fields are described by a perfect fluid of energy density $\rho$ and pressure $p$. In this case the relation between $\rho$ and $p$ is parametrized by an equation of state of the type $p=w\rho$, where $w$ is a constant. Therefore the conservation of matter stress-energy tensor ${T^{\mu\nu}}_{;\nu}=0$ with the metric 
\eqref{FLRWmetric} fixes the functional form 
\begin{equation}
\label{rhof}
\rho(a) = m \, a^{-3-3w}
\end{equation}

\noindent where $m$ is an arbitrary integration constant. It is now clear that $\mathcal{L}_m=-mNa^{-3w}$ \cite{2014greci}. One thus obtain the following Lagrangian without the York term \cite{1986York}
\begin{equation} \label{lag1}
\mathcal{L}=\, -\frac{3 \, a{\dot a }^{2}}{8\pi N(t)G(a)}+\frac{3 \, a N K}{8\pi G(a)} -\frac{ a^{3}N\Lambda(a)}{8\pi G(a)}-\frac{2 Nm}{a^{3w}} +\frac{3 \, a^{2}{\dot a }^{2}G'(a)}{8\pi N G(a)^{2}}+ \frac{d}{dt} \left(\frac{3 a^2{\dot a }^{2}}{8 \pi NG(a)}\right)
\end{equation}
where $G'(a)$ stands for the derivative of $G$ with respect to $a$. The York term \cite{1986York}, added as prescribed in \eqref{actionpunto}, cancels the total derivative.  

\section{Constraint Analysis}

The constraints analysis of this minisuperspace model is a lower dimensional application of the general  field theoretical analysis performed in the previous paragraphs. One gets a primary constraint 

\begin{equation}
p_N=\frac{\partial \mathcal{L}}{\partial {\dot N}}\approx 0\quad\mapsto\quad 
\phi_N(N,a,p_N,p_a)=p_N\approx 0 \;\;,
\label{primary}
\end{equation}

\noindent therefore one defines the canonocal Hamiltonian $H_{C}$ as 

\begin{equation}
H_C\equiv p_i q^i -\mathcal{L}|_{M}=p_a{\dot a}-\mathcal{L} \;\;.
\end{equation}

Here the momentum $p_a$ associated to the generalized coordinate $a(t)$ is given by

\begin{equation}
p_a\equiv\frac{\partial \mathcal{L}}{\partial {\dot a}}=-\frac{6 \, a\,{\dot a }}{8\pi N G(a)} \, (\eta(a)+1)
\end{equation}
where $\eta\equiv k\partial_k G_k$, with $k\sim a^{-1}$, defines the ``anomalous dimension'' of Newton's constant as a function of the scale factor
\begin{equation}
\label{anod}
\eta(a)=-\frac{a\,G'(a)}{G(a)} \;\;.
\end{equation}
The canonical Hamiltonian thus reads
\begin{equation}
H_C=-\frac{2 \pi NG(a)^2p^2_a}{3a(G(a)-aG'(a))}-\frac{3aNK}{8\pi G(a)}+\frac{a^3 \Lambda(a)N}{\pi G(a)}+\frac{2Nm}{a^{3w}}\;.
\label{Ham}
\end{equation}
Pursuing the Dirac's constraint analysis, one gets the Hamiltonian constarint as in \eqref{constraint}
\begin{equation}
\mathcal{H}=-\frac{2 \pi G(a)^2p^2_a}{3a(G(a)-aG'(a))}-\frac{3aK}{8\pi G(a)}+\frac{a^3 \Lambda(a)N}{\pi G(a)}+\frac{2m}{a^{3w}} \;\;.
\label{Hamconst}
\end{equation}
The total Hamiltonian $H_{T}$ is then
\begin{equation}
H_T=N\mathcal{H} +\lambda_N\phi_N
\label{total}
\end{equation}
Imposing the gauge $N=1$ as a constraint $N-1 \approx 0$, one has that $\phi_{N}$ becomes a second class constraint and $\lambda_{N}=0$, 
 \begin{eqnarray}
 \{N-1,\phi_{N}\}=1\;\;\;\;
\frac{d}{dt}(N-1)=\left\{N-1,H_T\right\}=\lambda_N=0\;\;.
\label{determino}
\end{eqnarray}

\section{Bouncing and Emergent Cosmologies from Asymptotic Safety}

The Hamiltonian constraint of the previous section provides the following RG-improved Friedmann equation

\begin{equation}
\frac{K}{a^2 H^2}-\frac{8\pi G(a)\, \rho+\Lambda(a)}{3 H^2}+\eta(a)+1=0 \;, \label{1fe}
\end{equation}

this implies an evolution equation for $a(t)$, provided $\eta(a) +1 \neq 0$

\begin{equation}
\dot{a}^2=-\tilde{V}_K(a)\equiv-\frac{K+V(a)}{\eta(a)+1}\;\;  {\textrm where}  \;\;
V(a)=\frac{{a}^2}{3}(8\pi G(a)\, \rho+\Lambda(a))
\label{evolvo}
\end{equation}
in which the scalings of $G(a)$ and $\Lambda(a)$ are determined by RG flow. It has been shown \cite{2000BonannoReuter} that under certain approximations the beta function for the Newton constant can be solved analytically and the beta function for the cosmological constant numerically. The running of the couplings in the early stages of the universe is completely captured by their behavour around the NGFP. One can approximates 

\begin{align}
&G(a)\simeq G_0 \left(1+G_0\, g_\ast^{-1}  a^{-2}\right)^{-1} \label{grun}\\ 
&\Lambda(a)\simeq \Lambda_0 + \lambda_\ast  a^{-2} \label{lrun} \;\;,
\end{align}

where equation \eqref{grun} is the expression of the running Newton's coupling constant in \cite{2000BonannoReuter} and \eqref{lrun} is the linearization of the beta function cosmological constant around the NGFP $(\lambda^{\star}, g^{\star})$.  $G_{0}$ and $\Lambda_{0}$ are the infrared value of the gravitation and cosmological constants and coincides the observed values. 

\begin{figure}
\begin{center}
\includegraphics[width=0.55\textwidth]{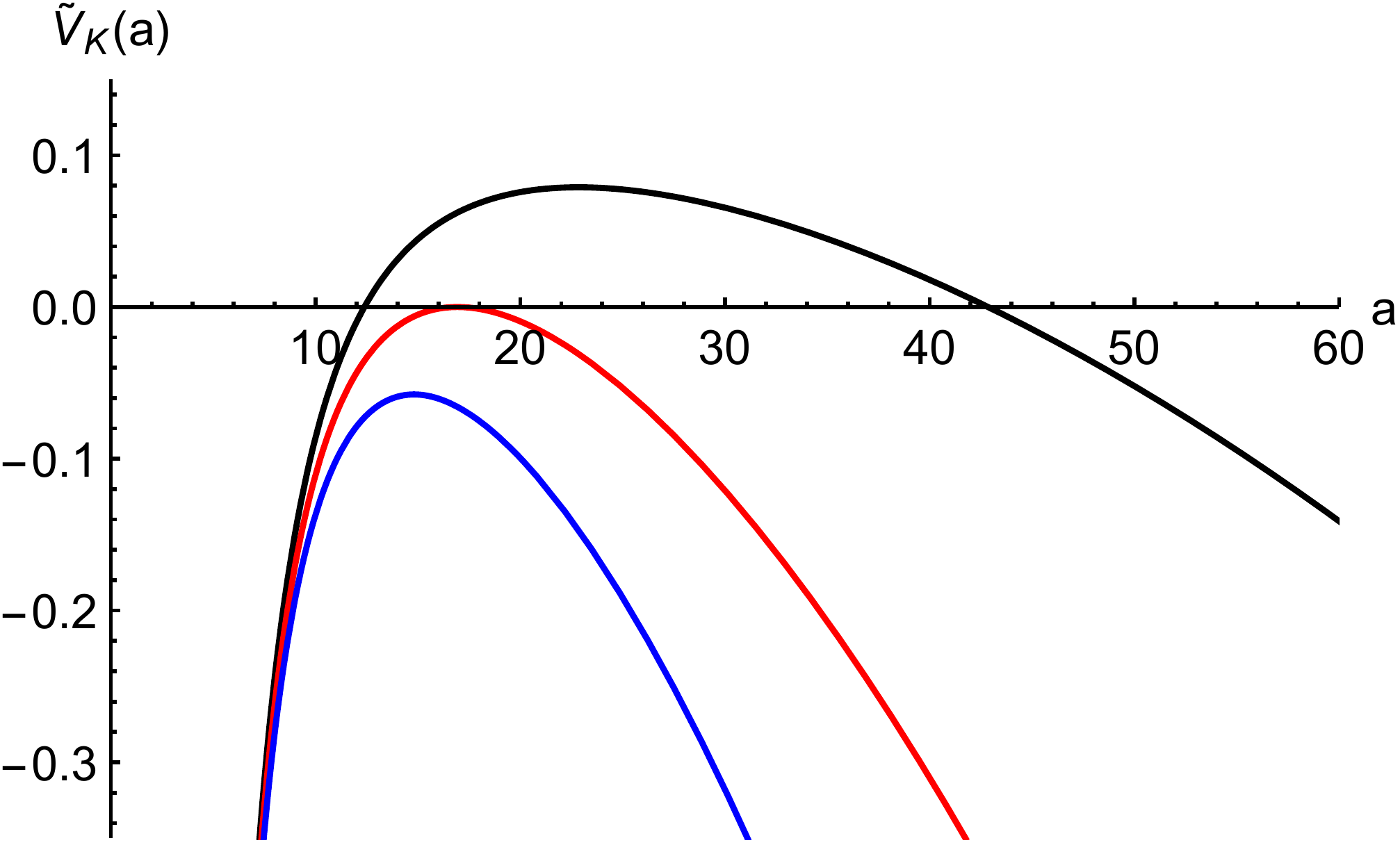}
\caption{The effective potential $\tilde{V}_K(a)$ for a
bouncing universe (black), emergent universe (red), singular universe (blue), for $K=0$, $w=1/3$, $g_\ast=0.1$, $\lambda_\ast=-0.5$ and $m=3$. Black, red and blue correspond to $\Lambda_0=2 \times 10^{-4}$, $\Lambda_0= 8.3 \times 10^{-4}$ and $\Lambda_0=1.5 \times 10^{-3}$ respectively.\label{fig1}}
\end{center}
\end{figure}

Cosmological implications of this analysis can be studied in the region $\tilde{V}_K(a)\leq 0$, as one may promptly check in equation \eqref{evolvo}. In particular, if $\tilde{V}_K(a)=0$ admits real solutions at some $a=a_b>0$, then $\tilde{V}_K(a)$ may give rise to either an emergent universe scenario or a bouncing model.
 The equation $\tilde{V}_K(a)=0$ implies
\begin{equation}
\left(a^2+\frac{G_0}{g_\ast}\right) \left(a^2+\frac{\lambda _\ast-3 K}{\Lambda_0}\right)+\frac{8\pi m\,G_0}{\Lambda_0}\,a^{1-3 w}
=0 \;\;. \label{poteqq}
\end{equation}
Bouncing solutions exist for some values of $w$. To simplify the discussion, one may restrict to the case of a radiation-dominated universe, $w=1/3$. Then eq.~\eqref{poteqq} has (at most) two solutions with non-negative real part. The number of such solutions determines the cosmological scenario arising from $\tilde{V}_K(a)$. In fact, no solutions implies no bounces and the universe has a singularity in the past, at $a=0$. This case corresponds to the blue line in Fig.~\ref{fig1}. On the other hand, a bouncing universe is realized when $\tilde{V}_K(a)$ has two different zeros. In particular if $\tilde{V}_K''(a)>0$ the scale factor oscillates between one minimum and one maximum value. On the contrary, if 
$\tilde{V}_K''(a)<0$, the universe has a bounce at either a minimum \textit{or} a maximum value of the scale factor (black line model in Fig. \ref{fig1}). Only in the former case the initial singularity is avoided.
The outcome of an emergent universe, from this model, represents the most interesting feature. It consists of \textit{past-eternal} inflationary phase which follows an initial quasi-static state. This universe starts at some minimum scale factor $a_{b}>0$. Later inflates and expands according to standard cosmology and the laws of General Relativity. The requirements to have an emergent universe at $a_{b}>0$ are $\tilde{V}_K''(a)<0$ and the double zero $\dot{a}_b=\ddot{a}_b=0$. This case is represented in the red line in Fig \ref{fig1}.
If one consider an early universe dominated by radiation ($w=\frac{1}{3}$), equation \eqref{poteqq} can be solved 
\begin{equation}
a_b^2 = -\frac{G_0 \Lambda_0 +g_\ast(\lambda_\ast-3K)}{2 g_\ast \Lambda_0} \pm\sqrt{\left(\frac{G_0 \Lambda_0-g_\ast(\lambda_\ast-3K)}{2 g_\ast \Lambda_0}\right)^2-\frac{8\pi m\,G_0}{\Lambda_0}} \;\;.
\end{equation}
Imposing that the previous equation has two coincident solutions (emergent universe condition) one determines the value of $m$. Furthermore $a_b^2$ has to be positive, which means 
\begin{equation}
 \label{cond}
\lambda_\ast-3K<-\frac{G_0\Lambda_0}{g_\ast} \;\;.
\end{equation}
In the classical case $\lambda_{*}=0$ ($g_{*}>0$) and then since the bare cosmological constant $\Lambda_{0}$ and the bare gravitational constant ${G_{0}}$ are positive, an emergent universe is possible for values of the spatial curvature $K>0$. In reference \cite{AlfiomeAlessia} it is discussed that the Asymptotic Safety Scenario is based on the evidence that there exists NGFP, such  that $\lambda_{*}\neq 0$. In particular there exists cases in which $\lambda_{*}$ is negative enough\cite{2017Alessia} \cite{AlfiomeAlessia} to allow also the cases $K=0$ and $K=-1$. 
Assuming that condition \eqref{cond} holds, in the case of an emergent universe eq. \eqref{evolvo} becomes 
\begin{equation}
\dot{a}^2=\frac{4 {g_\ast} a_b^2 \Lambda_0}{3 \left({g_\ast}a_b^2-{G_0} \right)} (a-a_b)^2
\end{equation}
in which the minimum scale factor $a_{b}$ is 
\begin{equation}
a_b=\sqrt{-\frac{G_0 \Lambda_0 +g_\ast(\lambda_\ast-3K)}{2 g_\ast \Lambda_0}} \;\;.
\end{equation}
Condition $\tilde{V}_K(a)\leq0$ implies 

\begin{equation}
a^2\geq a_b^2>\frac{G_0}{g_\ast}\;\;.
\end{equation}
(see also \cite{AlfiomeAlessia} for further details and discussions). 
One can investigate the behaviour of the early emergent universe close to $a_{b}$ linearizing the quantum equation \eqref{evolvo} around $a_{b}$. The approximate equation is then:
\begin{equation}
\dot{a}^2=\frac{4 {g_\ast} a_b^2 \Lambda_0}{3 \left({g_\ast}a_b^2-{G_0} \right)} (a-a_b)^2\;\;,
\end{equation}

then the general solution is 
\begin{equation}
a(t)=a_b+\epsilon\,\mathrm{exp}\left\{\sqrt{\frac{4 {g_\ast} a_b^2 \Lambda_0}{3 \left({g_\ast} a_b^2-{G_0}\right)}}\; t\right\} \;\;, 
\label{emergo}
\end{equation}

$\epsilon$ being an integration constant. It is evident that \eqref{emergo} exibits an emergent universe scenario with exponential evolution of the scale factor and then no need of a model with an  {\it{ad hoc}} inflation. The density parameter can be written 
\begin{equation}
\Omega-1=\frac{3 \left({g_\ast}a_b^2-{G_0} \right) K}{4 {g_\ast} a_b^4 \Lambda_0}\;e^{-2N_e}\;\;.
\end{equation}
The number $N_{e}$ of efolds is 

\begin{equation}
N_e\simeq\mathrm{log}\left(\frac{\epsilon}{a_b}\,\mathrm{exp}\left\{\sqrt{\frac{4 {g_\ast} a_b^2 \Lambda_0}{3 \left({g_\ast} a_b^2-{G_0}\right)}}\; t_e\right\}\right) \;\;,
\end{equation}
where $t_{e}$ is the cosmic time at the inflation exit. 

\section{Conclusions and Open Questions}

Hamiltonian (ADM) analysis of RG improved Einstein-Hilbert action with $G$ and $\Lambda$ as external, non geometrical field, has been performed. It has been showed that if one requires that this theory behaves like the Hamiltonian theory of Einstein General Relativity, that is the momentum constraints and the Hamiltonian constraint be the generators respectively of the space diffeomorphisms on $\Sigma$ and the time diffeomorphisms, then one cannot start from the ADM-metric \eqref{metricADM} but from ADM metric in Gaussian normal coordinates \eqref{normalcoordinates}. 

An immediate application of the above considerations is FLRW cosmology  in the minisupersapce approach using Dirac's constraint analysis. It  generates sub-Planckian cosmological models via Asymptotic Safety.  They exhibit bouncing and emergent Universes. The latter ones are solution of the equations of motion  also in cases $K=-1,0$, that are impossible to draw from classical General Relativity. 

Although this analysis shows that RG improved Einstein-Hilbert action with $G$ and $\Lambda$ as external fields can be cast in the Hamiltonian formalism only in the case of ADM metric in Gaussian normal coordinates, one can still legitimately ask if there exists cases and/or particular foliations in which one does not need to loose space diffeomorphisms in order to make sense of the Hamiltonian formalism. In order to throw light on this issue, it could be useful, following the suggestions of section $2$, to study the Hamiltonian formalism of the Branse-Dicke theory.

In the same direction ADM formalism for Black Holes could result quite enlightening . Here a completely different symmetry implies a different ADM foliation, which could, eventually, help to answer previous questions. 

\section{Acknowledgment}

G. Gionti thanks the organizers of the Corf\'u Summer Workshop on Testing the Fundamental Physics Principles. He  is grateful to A. Bonanno and A. Platania for stimulating discussions and collaboration on this work. He has also enjoyed the hospitality at Osservatorio Astrofisico di Catania while part of this research has been carried out.   

\bibliographystyle{JHEP}
\bibliography{hamiltonian}

\providecommand{\href}[2]{#2}\begingroup\raggedright\begin{thebibliography}{10}

\bibitem{lauscherreuter2001}
O.~{Lauscher} and M.~{Reuter}, \emph{{Is quantum Einstein gravity
  nonperturbatively renormalizable?}},
  \href{https://doi.org/10.1088/0264-9381/19/3/304}{\emph{Classical and Quantum
  Gravity} {\bfseries 19} (2002) 483}
  [\href{https://arxiv.org/abs/hep-th/0110021}{{\ttfamily hep-th/0110021}}].

\bibitem{wilsonfourQFT}
K.~G. Wilson, \emph{{Quantum field theory models in less than
  four-dimensions}}, \href{https://doi.org/10.1103/PhysRevD.7.2911}{\emph{Phys.
  Rev.} {\bfseries D7} (1973) 2911}.

\bibitem{NiederReuter}
M.~{Niedermaier} and M.~{Reuter}, \emph{{The Asymptotic Safety Scenario in
  Quantum Gravity}}, \href{https://doi.org/10.12942/lrr-2006-5}{\emph{Living
  Reviews in Relativity} {\bfseries 9} (2006) 5}.

\bibitem{1979weinberg}
S.~{Weinberg}, \emph{{Ultraviolet divergences in quantum theories of
  gravitation.}},  in \emph{General Relativity: An Einstein centenary survey}
  (S.~W. {Hawking} and W.~{Israel}, eds.), pp.~790--831, 1979.

\bibitem{2001ReuterSaueressig}
M.~Reuter and F.~Saueressig, \emph{{Renormalization group flow of quantum
  gravity in the Einstein-Hilbert truncation}},
  \href{https://doi.org/10.1103/PhysRevD.65.065016}{\emph{Phys. Rev.}
  {\bfseries D65} (2002) 065016}
  [\href{https://arxiv.org/abs/hep-th/0110054}{{\ttfamily hep-th/0110054}}].

\bibitem{Guarnieriphd}
F.~Guarnieri, \emph{{Renormalization group flow of scalar models in gravity}},
  Master's thesis, Rome U., 2014-04-08.

\bibitem{WetterichFRG}
J.~Berges, N.~Tetradis and C.~Wetterich, \emph{{Nonperturbative renormalization
  flow in quantum field theory and statistical physics}},
  \href{https://doi.org/10.1016/S0370-1573(01)00098-9}{\emph{Phys. Rept.}
  {\bfseries 363} (2002) 223}
  [\href{https://arxiv.org/abs/hep-ph/0005122}{{\ttfamily hep-ph/0005122}}].

\bibitem{1972Dirac}
P.~A.~M. Dirac, \emph{{Consequences of Varying G}},
  \href{https://doi.org/10.1063/1.31597}{\emph{AIP Conf. Proc.} {\bfseries 48}
  (1978) 169}.

\bibitem{BransDicke}
C.~{Brans} and R.~H. {Dicke}, \emph{{Mach's Principle and a Relativistic Theory
  of Gravitation}},
  \href{https://doi.org/10.1103/PhysRev.124.925}{\emph{Physical Review}
  {\bfseries 124} (1961) 925}.

\bibitem{ReuterWeyer}
M.~Reuter and H.~Weyer, \emph{{Renormalization group improved gravitational
  actions: A Brans-Dicke approach}},
  \href{https://doi.org/10.1103/PhysRevD.69.104022}{\emph{Phys. Rev.}
  {\bfseries D69} (2004) 104022}
  [\href{https://arxiv.org/abs/hep-th/0311196}{{\ttfamily hep-th/0311196}}].

\bibitem{Manrique}
E.~{Manrique}, S.~{Rechenberger} and F.~{Saueressig}, \emph{{Asymptotically
  Safe Lorentzian Gravity}},
  \href{https://doi.org/10.1103/PhysRevLett.106.251302}{\emph{Physical Review
  Letters} {\bfseries 106} (2011) 251302}
  [\href{https://arxiv.org/abs/1102.5012}{{\ttfamily 1102.5012}}].

\bibitem{menotti}
P.~{Menotti}, \emph{{Lectures on gravitation}}, {\emph{ArXiv e-prints} (2017) }
  [\href{https://arxiv.org/abs/1703.05155}{{\ttfamily 1703.05155}}].

\bibitem{AlfioGiampieroRubano}
A.~{Bonanno}, G.~{Esposito} and C.~{Rubano}, \emph{{Arnowitt Deser Misner
  gravity with variable G and {$\Lambda$} and fixed-point cosmologies from the
  renormalization group}},
  \href{https://doi.org/10.1088/0264-9381/21/21/017}{\emph{Classical and
  Quantum Gravity} {\bfseries 21} (2004) 5005}
  [\href{https://arxiv.org/abs/gr-qc/0403115}{{\ttfamily gr-qc/0403115}}].

\bibitem{espositolibro}
G.~Esposito, \emph{{Quantum gravity, quantum cosmology and Lorentzian
  geometries}}, \href{https://doi.org/10.1007/978-3-540-47295-7}{\emph{Lect.
  Notes Phys. Monogr.} {\bfseries 12} (1992) 1}.

\bibitem{Arnoldlibro}
V.~I. {Arnold}, \emph{{Mathematical methods of classical mechanics}}. Graduate
  texts in mathematics, New York: Springer, 1978.

\bibitem{dirac1966}
P.~A.~M. Dirac, \emph{Lectures on quantum field theory}. Yeshiva Univ., 1966.

\bibitem{Wiltshire}
D.~L. Wiltshire, \emph{{An Introduction to quantum cosmology}},  in
  \emph{{Cosmology: The Physics of the Universe. Proceedings, 8th Physics
  Summer School, Canberra, Australia, Jan 16-Feb 3, 1995}}, pp.~473--531, 1995,
  \href{https://arxiv.org/abs/gr-qc/0101003}{{\ttfamily gr-qc/0101003}}.

\bibitem{AlfiomeAlessia}
A.~{Bonanno}, S.~J.~G. {Gionti} and A.~{Platania}, \emph{{Bouncing and emergent
  cosmologies from Arnowitt Deser Misner RG flows}},
  \href{https://doi.org/10.1088/1361-6382/aaa535}{\emph{Classical and Quantum
  Gravity} {\bfseries 35} (2018) 065004}
  [\href{https://arxiv.org/abs/1710.06317}{{\ttfamily 1710.06317}}].

\bibitem{2002PhRDAlfioReuter}
A.~Bonanno and M.~Reuter, \emph{{Cosmology of the Planck era from a
  renormalization group for quantum gravity}},
  \href{https://doi.org/10.1103/PhysRevD.65.043508}{\emph{Phys. Rev.}
  {\bfseries D65} (2002) 043508}
  [\href{https://arxiv.org/abs/hep-th/0106133}{{\ttfamily hep-th/0106133}}].

\bibitem{2002PhLBAlfioReuter}
A.~{Bonanno} and M.~{Reuter}, \emph{{Cosmology with self-adjusting vacuum
  energy density from a renormalization group fixed point}},
  \href{https://doi.org/10.1016/S0370-2693(01)01522-2}{\emph{Physics Letters B}
  {\bfseries 527} (2002) 9}
  [\href{https://arxiv.org/abs/astro-ph/0106468}{{\ttfamily
  astro-ph/0106468}}].

\bibitem{1960JMPADM}
R.~Arnowitt, S.~Deser and C.~W. Misner, \emph{Canonical variables for general
  relativity}, \href{https://doi.org/10.1103/PhysRev.117.1595}{\emph{Phys.
  Rev.} {\bfseries 117} (1960) 1595}.

\bibitem{1961PhRDADM}
R.~Arnowitt, S.~Deser and C.~W. Misner, \emph{Wave zone in general relativity},
  \href{https://doi.org/10.1103/PhysRev.121.1556}{\emph{Phys. Rev.} {\bfseries
  121} (1961) 1556}.

\bibitem{1960PhR1ADM}
R.~Arnowitt, S.~Deser and C.~W. Misner, \emph{Energy and the criteria for
  radiation in general relativity},
  \href{https://doi.org/10.1103/PhysRev.118.1100}{\emph{Phys. Rev.} {\bfseries
  118} (1960) 1100}.

\bibitem{1960PhR2ADM}
R.~Arnowitt, S.~Deser and C.~W. Misner, \emph{Gravitational-electromagnetic
  coupling and the classical self-energy problem},
  \href{https://doi.org/10.1103/PhysRev.120.313}{\emph{Phys. Rev.} {\bfseries
  120} (1960) 313}.

\bibitem{2014greci}
N.~{Dimakis}, T.~{Christodoulakis} and P.~A. {Terzis}, \emph{{FLRW metric f(R)
  cosmology with a perfect fluid by generating integrals of motion}},
  \href{https://doi.org/10.1016/j.geomphys.2013.12.001}{\emph{Journal of
  Geometry and Physics} {\bfseries 77} (2014) 97}
  [\href{https://arxiv.org/abs/1311.4358}{{\ttfamily 1311.4358}}].

\bibitem{1986York}
J.~W. {York}, \emph{{Boundary terms in the action principles of general
  relativity}}, \href{https://doi.org/10.1007/BF01889475}{\emph{Foundations of
  Physics} {\bfseries 16} (1986) 249}.

\bibitem{2000BonannoReuter}
A.~Bonanno and M.~Reuter, \emph{{Renormalization group improved black hole
  space-times}}, \href{https://doi.org/10.1103/PhysRevD.62.043008}{\emph{Phys.
  Rev.} {\bfseries D62} (2000) 043008}
  [\href{https://arxiv.org/abs/hep-th/0002196}{{\ttfamily hep-th/0002196}}].

\bibitem{2017Alessia}
J.~Biemans, A.~Platania and F.~Saueressig, \emph{{Renormalization group fixed
  points of foliated gravity-matter systems}},
  \href{https://doi.org/10.1007/JHEP05(2017)093}{\emph{JHEP} {\bfseries 05}
  (2017) 093} [\href{https://arxiv.org/abs/1702.06539}{{\ttfamily
  1702.06539}}].

\end{thebibliography}\endgroup

\end{document}